\documentclass[final,5p,times,twocolumn]{elsarticle}

\usepackage{amssymb}
\usepackage{graphicx}
\usepackage{float}
\usepackage{amsmath}
\usepackage{algorithm}
\usepackage{algpseudocode}
\usepackage{booktabs}
\usepackage{hyperref}
\usepackage{array}
\usepackage{orcidlink}

\newcommand{\leadershipWeight}{\texttt{leadership\_weight}}
\newcommand{\sort}{\texttt{sort}}

\journal{.}

\begin{document}

\begin{frontmatter}



\title{Research impact evaluation based on effective authorship contribution sensitivity: h-leadership index }

\author[inst1]{Hardik A. Jain\orcidlink{0000-0002-4129-3183}}
\author[inst2]{Rohitash Chandra\orcidlink{0000-0001-6353-1464}}

\affiliation[inst1]{organization={School of Electrical Engineering, Computing and Mathematical Sciences},
            addressline={Curtin University},
            city={Perth}, 
            country={Australia}}

\affiliation[inst2]{organization={Transitional Artificial Intelligence Research Group, School of Mathematics and Statistics},
            addressline={University of New South Wales},
            city={Sydney}, 
            country={Australia}}


\begin{abstract}
The evaluation of a researcher's performance has traditionally relied on various bibliometric measures, with the h-index being one of the most prominent. However, the h-index only accounts for the number of citations received in a publication and does not account for other factors, such as the number of authors or their specific contributions in collaborative works. Therefore, the h-index has been placed under scrutiny as it has motivated academic integrity issues where non-contributing authors get authorship merely for raising their h-index. In this study, we comprehensively evaluate existing metrics' ability to account for authorship contribution by their position and introduce a novel h-index variant, the h-leadership index. The h-leadership index aims to advance the fair evaluation of academic contributions in multi-authored publications by giving importance to authorship position beyond the first and last authors, focused by Stanford's ranking of the top 2 \% of world scientists. We assign weighted citations based on a modified complementary unit Gaussian curve, ensuring that the contributions of middle authors are appropriately recognised. We apply the h-leadership index to analyse the top 50 researchers across the Group of 8 (Go8) universities in Australia, demonstrating its potential to provide a more balanced assessment of research performance. We provide open-source software to extend the work further.
\end{abstract}


\end{frontmatter}
 
\section{Introduction}

The evaluation of a researcher's performance has used various bibliometric measures based mainly on citations, with the h-index being one of the most prominent. Hirsch introduced the h-index in 2005~\cite{hirsch2005index}, which combines a list of publications by a researcher along with the number of citations. The h-index is simple to compute and easy to understand and can be used for evaluating a school, institution, and journal profile, apart from individual researchers. The Google Scholar Metrics \footnote{Google Scholar Metrics: \url{https://scholar.google.com/citations?view_op=top_venues}} uses the h-index to rank academic journals \cite{mingers2017evaluating,falagas2008comparison} and provides insights on universities and individual researcher profiles. It has been argued that Google Scholar Metrics can be easily manipulated with poor citation culture and academic dishonesty \cite{delgado2012google,lopez2012manipulating}. Therefore, the h-index has several limitations~\cite{costas2007h, glanzel2006opportunities, haustein2014use} with the primary criticism being its insensitivity to highly cited papers. Once a paper is included in the h-core (the set of papers that contribute to the h-index), additional citations do not affect the index~\cite{schreiber2008empirical}. This means that even if a paper receives a significant number of citations, the h-index does not reflect this increase, potentially underestimating the impact of influential works. Furthermore, the h-index exhibits a low degree of discrimination, making it less effective in differentiating between researchers with similar citation patterns since authorship contributions (positions) are not taken into account~\cite{alonso2009h}. Another limitation is that the h-index can only increase over time, allowing scientists to rest on their past achievements without accounting for new citations, which may not accurately represent their current research impact~\cite{alonso2009h}; therefore, Google Scholar includes the h5-index, which looks at citations received in the last five years. Finally, the h-index does not account for the varying contribution of co-authors in multi-authored papers, which can skew the perceived impact of collaborative research efforts~\cite{schreiber2008modification}.


Egghe proposed the g-index in 2006 to address some of the limitations of the h-index by giving more weight to highly cited $g$ papers (g-core) such that together they contribute at least $g^2$ citations~\cite{egghe2006improvement, egghe2006theory}. However, similar to the h-index, the g-index neglects papers and citations outside the g-core, focusing only on the subset of papers that contribute to the index. This limitation can result in an incomplete assessment of a researcher's overall impact. Additionally, the g-index can reach a saturation point where subsequent citations do not influence the index unless new papers are published. Jin ~\cite{jin2007r} introduced the AR-index in 2007, considering the age of publications and giving more weight to recent works that highlight current research activities. It ignores the impact of h-tail papers (papers not in the h-core) and their citations. This can lead to underestimating the research impact for many authors whose influential works lie outside the h-core. Moreover, the AR index shares some of the h-index's limitations, such as low discrimination ability and insensitivity to highly cited papers. The AR-index attempts to provide a temporal dimension to citation analysis but still falls short in addressing the complexities of authorship positions and collaborative efforts. This prompted the development of metrics such as the hm-index~\cite{schreiber2008modification} and h-frac index~\cite{koltun2021h} that utilise weighted citations to provide a more nuanced assessment by considering the number of contributing authors. Recently, more complex measures such as c-score have emerged that provide a composite score combining six factors, namely citations, h-index, hm-index, single, first and last authorship percentages~\cite{ioannidis2019standardized}. The c-score has gained prominence, ranking top authors by factoring in multiple dimensions of their contributions, leading to a more standardised ranking system. An updated version of c-score rankings gets released annually, known as Stanford University's Top 2\% Scientists list\footnote{\url{https://elsevier.digitalcommonsdata.com/datasets/btchxktzyw/7}}. However, despite its robustness, the c-score and similar metrics fall short by inadequately considering the authorship positions beyond the first and last authors. Furthermore, these lists do not apply to authorship culture, where surnames are used to list the authors alphabetically.

The bibliometrics for evaluating academic performance has become increasingly prevalent in recent decades to assess the impact and influence of a university and researcher's profile. However, the application of bibliometrics involves multiple complexities and loopholes that are easy to exploit, leading to its potential misuse~\cite{gingras2016bibliometrics}. Therefore, we first explore the development, benefits, and challenges associated with popular bibliometrics, as well as the introduction of composite indicators to provide a more nuanced evaluation of scientific impact. In our study, we introduce a new way to measure research contributions called the h-leadership index. This index assigns credit based on the position of the author such that the first and last authors of a paper get the most credit, the second and second-last authors slightly less, and so on, based on a modified complementary unit Gaussian distribution. The h-leadership index assigns weights to an author's citation for a work based on their authorship position, ensuring the authorship position is accounted for along with the number of citations and number of publications being measured. This new metric thus acknowledges and quantifies the diverse contributions of researchers within collaborative environments. We also present and compare the new metrics with more widely used bibliometrics such as h-index~\cite{hirsch2005index}, hm-index~\cite{schreiber2008modification} and c-score~\cite{ioannidis2019standardized}.

\begin{figure}
    \centering 
    \includegraphics[width=1\linewidth] {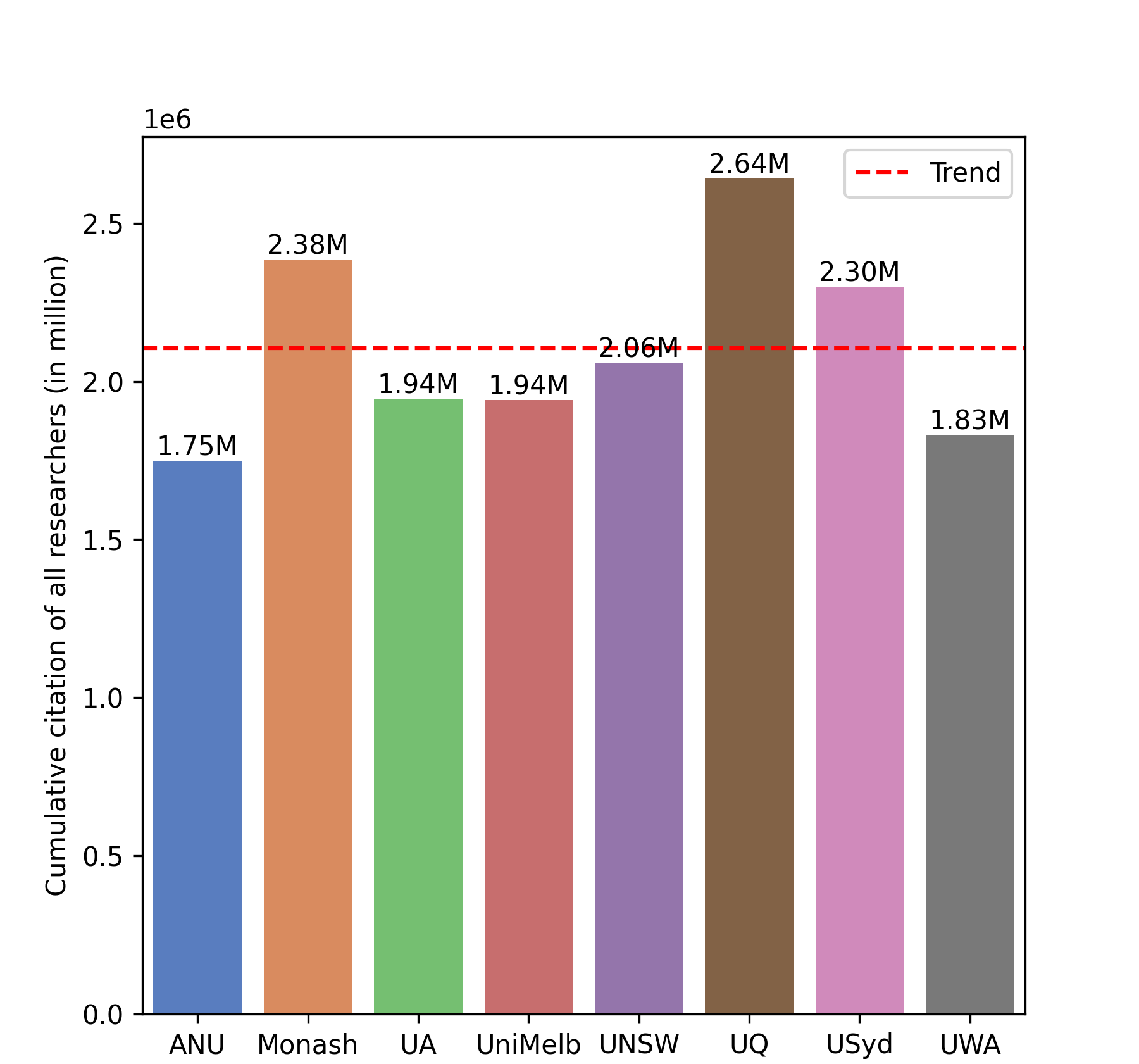}
    \caption{Cumulative citations received by Go8 researchers.}
    \label{fig:cites}
\end{figure}

As a case study to demonstrate the effectiveness of the h-leadership index, we use Australia's group of eight (Go8) institutions with retrieved data from the top 50 affiliated researchers. Data was retrieved using Scopus APIs\footnote{Elsevier Developer Portal: \url{https://dev.elsevier.com/}}, which offer generous limits (up to 5000 requests/weekly) for academic research. Go8 universities showcase very high research output with a mean citation of 2.1~million as shown in Figure~\ref{fig:cites}. These institutions are renowned for their substantial contributions to research across various disciplines, making them ideal subjects for examining advanced bibliometric methodologies.

The rest of the paper is organised as follows. Section~\ref{sec:literature} provides an overview of the related metrics under consideration, and Section~\ref{sec:methodology} outlines the methodology employed to identify and collect data. Section~\ref{sec:results} presents the results of metric computations and their implications for research evaluation practices. Finally, Section~\ref{sec:conclusion} discusses the broader implications of the findings and suggests avenues for future research in bibliometric methodologies.

\section{Related Work}\label{sec:literature}

\subsection{h-index and hm-index}


The \textit{h-index} provides a balanced measure by considering both the quantity and the quality of publications. A researcher has an h-index of $h$ if $h$ of their papers have received at least $h$ citations each \cite{bornmann2007we}. Formally, the h-index is the largest number $h$ such that at least $h$ papers have been cited $h$ or more times. It can be expressed mathematically as:

\begin{equation}
    h = \max_r (r \mid r \leq c(r))
    \tag{1}
\end{equation}
where $r$ is the rank of the paper when sorted by the number of citations $c(r)$~\cite{schreiber2008share}.


The h-index is now widely used in academia for assessing the impact of researchers, departments, and institutions~\citep{lazaridis2010ranking, svider2013use}. It balances publication quantity with citation quality, making it useful in hiring, funding, and promotion decisions \cite{wang2022using} and used with variants \cite{bornmann2008there}.


The \textit{hm-index}, or \textit{modified h-index}, was proposed as an adjustment to the h-index to address some of its inherent limitations, particularly the equal weighting of all co-authors in multi-authored papers. The hm-index aims to account for the contribution of each co-author by normalising the citation count based on the number of authors. The hm-index is calculated by adjusting the number of citations for each paper according to the number of authors, thus providing a more equitable measure for individual contribution. It is defined as follows: a researcher has an hm-index of $h_m$ if $h_m$ of their publications have an adjusted citation count (based on the number of authors) that is at least $h_m$ \cite{schreiber2008modification}. This accounts for varying levels of contribution across papers, particularly in fields where large collaborations are common. It can be expressed mathematically as:

\begin{equation}
  h_m = \max_r \left(r \mid \sum_{r'=1}^r \frac{1}{a(r')} \leq c(r) \right)
  \tag{2}
\end{equation}
where $a(r)$ is the number of authors of the $r$-th paper~\cite{schreiber2008share}.


The hm-index is particularly useful in disciplines such as physics \cite{tietze2019h}, biomedical research, and large-scale collaborations, where author lists can be extensive. Although the hm-index improves upon the traditional h-index by addressing the issue of multi-author papers, it still shares some limitations, such as being sensitive to the citation practices of different fields~\cite{bornmann2020should} and not distinguishing between positive and negative citations~\cite{budi2023understanding}.

\subsection{Composite Indicators}

A composite indicator (c-score) has been proposed and developed by Ioannidis et al. \cite{ioannidis2019standardized} to address some of the limitations of h-based indexes. It includes six key metrics: total citations, h-index, co-authorship-adjusted hm-index, number of citations to papers as single author, number of citations to papers as single or first author, and number of citations to papers as single, first, or last author. The composite indicator can be expressed as:

\begin{equation}
C_{\text{score}} = \sum_{i=1}^{6} \frac{\log(1+C_i)}{\max\log(1+C_i)}
\tag{3}
\end{equation}

where $C_i$ represents the six metrics: total citations, h-index, $h_m$-index, number of single-authored papers, number of single or first-authored papers, and number of single, first, or last-authored papers. The c-score offers several advantages over single-metric evaluations by mitigating the impact of field-specific citation practices by incorporating diverse metrics that capture different aspects of a researcher's contributions. Additionally, it provides separate data for career-long and single-year impacts, allowing for more dynamic and temporally relevant assessments. 

\section{Methodology}\label{sec:methodology}

We carried out the study in four steps: i) review of existing metrics, ii) design of the h-leadership index, iii) data retrieval from scopus and iv) analysis of the research impact. 

\subsection{Framework for the h-leadership index}\label{sec:h-leadership}

The h-leadership index is a novel metric designed to address the limitations of existing bibliometric measures, particularly in multi-authored publications. Traditional metrics like the h-index, while effective in measuring citation impact, often fail to account for the varying contributions of authors based on their positions within a paper. The h-leadership index mitigates this by introducing a weighted citation scheme that allocates points to authors based on their position. The first and last few authors receive the highest weights, reflecting their significant contributions, while middle authors receive slightly less. The weighting scheme is based on a modified complementary unit Gaussian curve, which ensures a balanced recognition of contributions across all authorship positions.

We note that the Gaussian distribution is a symmetric bell-shaped curve centred around the mean, used to represent a normal distribution. Furthermore, a complementary Gaussian distribution is defined as $$1-\texttt{Gaussian}(x)$$, where the peak is at the ends rather than in the middle. This transformation effectively inverts the bell curve, giving higher values to the tails (i.e., positions at the start and end). The lowest value in complementary Gaussian distribution is 0, which can leave some authors with no contribution. Thus, we modified the complementary unit Gaussian distribution as:

\begin{equation}
    w(x) = \begin{cases}
        0.3 + 0.7*\left(1 - \frac{1}{\sigma \sqrt{2\pi}} e^{-\frac{(x-\mu)^2}{2\sigma^2}}\right) & \text{if } x \leq \mu~\text{or}~n-x \leq \mu \\
        0.3 & \text{otherwise}
    \end{cases}
    \tag{4}
\end{equation}

where, $n$ is the total number of authors in a publication, $x$ is the given author's position, $\mu$ is the mean (set to 50), and $\sigma$ is the standard deviation.

\begin{figure}
    \centering
    \includegraphics[width=1\linewidth]{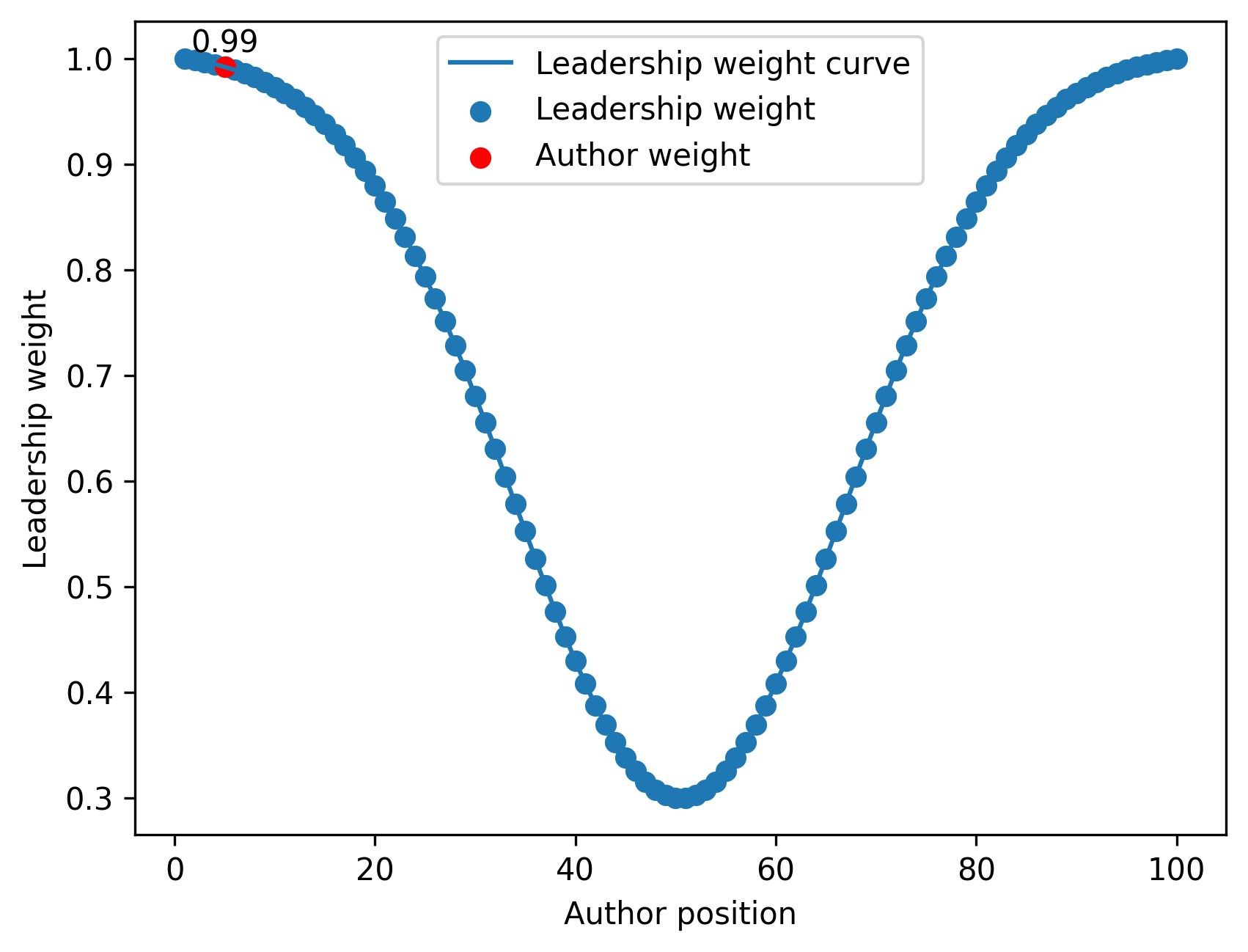}
    \caption{Variation of weights assigned to authors based on their authorship position in the h-leadership index. First and last authorship positions get the highest contribution at 1, with a marginal decrement as the authorship position moves towards the middle authors. The leadership weight assigned at position 5 is 0.992, which reaches 0.3 by position 50. If the position from the first and last authors is greater than 50, then the weight stays static at 0.3 (lowest assigned).}
    \label{fig:h-leadership}
\end{figure}

As shown in Figure~\ref{fig:h-leadership}, the weights are assigned gradually, with the contribution of the 5th position still being given the same weight as 99 percent of the first author. The minimum weight assigned is 0.3 for any position beyond the 50th author if authorship exceeds 100. Due to symmetry, the last set of authors are given the same contribution as the first set of author positions. The weights do not vary based on the number of authors but only based on the authorship position so that studies having many authors are not unnecessarily penalised for having a larger collaboration. Unlike traditional metrics that often treat all authors equally or highlight only first/last positions, the h-leadership index ensures that the contributions of all authors are measured based on their position.

The choice of a modified complementary unit Gaussian curve for weighting is justified as it naturally emphasises the first and last authors, who typically lead and finalise the research while reducing the weight for middle authors who may have had more supportive roles. Other potential models, such as linear or exponential decay functions, were considered but found to either overly penalise or under-represent certain authorship positions, leading to less accurate assessments. The modified complementary unit Gaussian model offers the best balance, accurately reflecting the typical contributions across different authorship positions.

Using the modified complementary unit Gaussian distribution, we calculate the h-leadership index by first assigning weights to citations based on the author's position in the author list. We then sort the list of weighted citations in descending order and then determine the h-leadership index as the largest number $h$ such that the author has $h$ papers, each with at least $h$ weighted citations. Mathematically, let $C_i$ be the citations received by the $i$-th paper, and $w(x)$ be the weight assigned based on the author's position $x$. The h-leadership index $h_L$ is then given by:

\begin{equation}
    h_L = \max \{ h : \text{at least } h \text{ papers have } C_i \times w(x) \geq h \}
    \tag{5}
\end{equation}

Algorithm~\ref{alg:h_leadership_index} calculates the h-leadership index for a given author, identified by their \texttt{scopus\_id}, based on a list of their \texttt{publications}. The algorithm works by iterating through each publication to compute weighted citations for the author. It then computes the weight based on the author's position in the list of co-authors, using a function \texttt{leadership\_weight(x: author position, n: number of authors in a publication)}. The citations are accumulated and sorted in descending order, and the h-leadership index is determined as the maximum number of publications with at least $h$ citations, adjusted by the author's position in the author list. 

\begin{algorithm}
\begin{algorithmic}[1]
\caption{Calculation of h-Leadership Index}
\label{alg:h_leadership_index}
\State \textbf{Input:} $scopus\_id$, $publications$
\State \textbf{Initialize:} $cum\_l\_weight \gets 0$, $weighted\_citations \gets []$
\For {each $pub$ in $publications$}
    \State $authors \gets$ list of $scopus\_id$ from $pub$
    \If {$pub.citations = 0$}
        \State $weighted\_citations \gets weighted\_citations \cup [0]$
    \ElsIf {$scopus\_id \in authors$}
        \State $author\_position \gets$ index of $scopus\_id$ in $authors$
        \State $l\_weight \gets \leadershipWeight(author\_position, n=|authors|)$
        \State $cum\_l\_weight \gets cum\_l\_weight + l\_weight$
        \State $weighted\_citations \gets weighted\_citations \cup [pub.citations \times l\_weight]$
    \EndIf
\EndFor
\State $\sort(weighted\_citations, \text{descending})$
\State \textbf{Initialize:} $h\_leadership \gets 0$
\For {$i = 1$ to $|weighted\_citations|$}
    \If {$weighted\_citations[i] \geq i$}
        \State $h\_leadership \gets i$
    \Else
        \State \textbf{break}
    \EndIf
\EndFor
\State \textbf{Return:} $h\_leadership$
\end{algorithmic}
\end{algorithm}

\subsection{Data retrieval}\label{subsec:data}

\begin{figure*}
    \centering
    \includegraphics[width=0.75\linewidth]{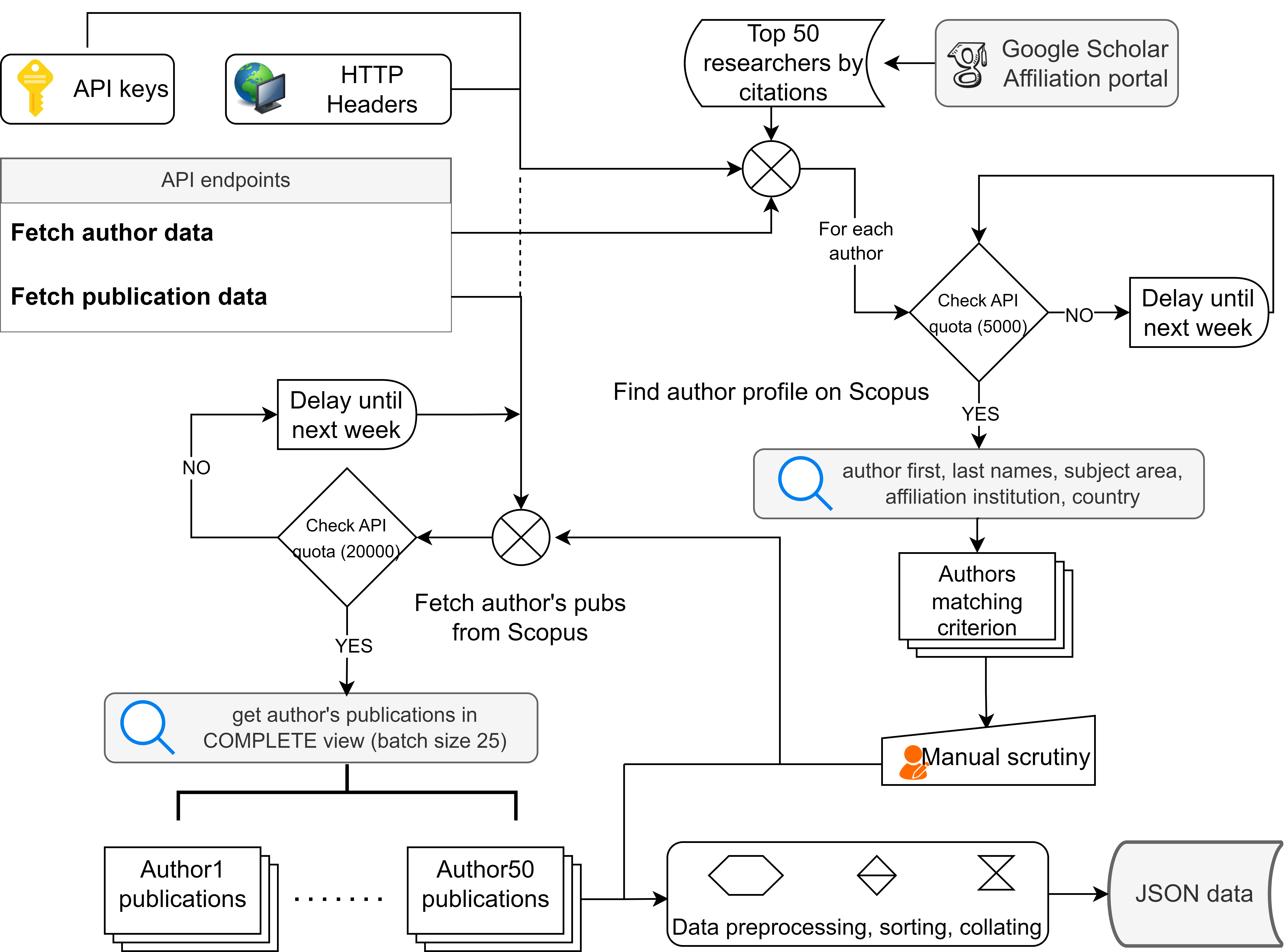}
    \caption{Automated pipeline created in Python for data retrieval from Scopus APIs with sufficient manual scrutiny performed to ensure data quality and reliability. Fetch author data and fetch publication data refer to endpoints https://api.elsevier.com/content/search/author and https://api.elsevier.com/content/search/scopus respectively.}
    \label{fig:data_retrieval}
\end{figure*}

We retrieved the data for this using Scopus, a comprehensive database for academic research outputs, as shown in Algorithm~\ref{alg:fetch}. The focus was on retrieving publication records for the top 50 researchers based on citations affiliated with Go8. We identified the top researchers at these universities using their Google Scholar affiliation page and a list of researchers from these universities. The need for using Google Scholar in determining the top researchers arose from the observation that Scopus sometimes list scholars in the same university as affiliated with their school instead of the university, making it difficult to group them. Our approach ensures a robust dataset encompassing a diverse range of publication records and accurate citation metrics essential for computing the proposed h-leadership index. Using the Scopus API, we searched for authors based on their names, affiliations, and countries. For each identified author, we extracted publication meta-data, including titles, authorship details, citation counts, publication venues, and other relevant metadata. The retrieved data was stored in a structured format, enabling subsequent analysis. We also compiled the list of the top domestic and international collaborators for these researchers to analyse the overall networking that pertains to modern research studies. The output of this process is a list of correctly matched researchers for each affiliation, along with their complete publication history, obtained through iterative requests from the Scopus API as shown in Figure~\ref{fig:data_retrieval}.
We selected Scopus as the primary data source due to its extensive coverage of scholarly literature and reliable citation-tracking capabilities. Despite challenges related to affiliation consistency, Scopus remains integral to bibliometric research, offering access to a vast database crucial for conducting comprehensive analyses of publication impact and authorship dynamics~\cite{donthu2021conduct}.

\begin{algorithm}[H]
\caption{Extraction of Research Profiles and Publications by Affiliation}
\label{alg:fetch}
\begin{algorithmic}[1]
\State \textbf{Input:} List of affiliations with their corresponding list of researcher names
\State \textbf{Output:} List of researchers and their publications

\For{each affiliation in affiliations}
    \State affiliation\_id $\gets$ affiliation.id
    \State researcher\_list $\gets$ affiliation.researcher\_names
    \State matched\_researchers $\gets$ \textbf{fetch\_authors\_by\_affiliation}(affiliation\_id, researcher\_list)
    
    \For{each researcher in matched\_researchers}
        \If{researcher.subject\_area == target\_subject\_area \textbf{and} 
            researcher.document\_count $\geq$ min\_document\_count \textbf{and}
            researcher.country == target\_country \textbf{and} 
            researcher.city == target\_city}
            \State correct\_author $\gets$ researcher
            \State scopus\_id $\gets$ correct\_author.scopus\_id
            \State publications $\gets$ []
            \State start $\gets$ 0
            \State count $\gets$ 25
            
            \Repeat
                \State response $\gets$ \textbf{scopus\_search\_API}(scopus\_id, start, count)
                \State publications.append(response.publications)
                \State start $\gets$ start + count
            \Until{response.has\_more == \textbf{false}}
            
            \State \textbf{store} publications for correct\_author
        \EndIf
    \EndFor
\EndFor
\end{algorithmic}
\end{algorithm}

We retrieved a total of 400 researcher records (50 each from the eight institutions) with information on their Scopus ID (identifier), name, publications, affiliation, city, country, number of documents published, and subject areas. Each publication includes information on the article title, co-authors, citations, publication name, venue, cover date, ISSN (International Standard Serial Number), issue, page range and DOI (Digital Object Identifier). The co-authors' affiliations are also listed.

\subsection{Technical details}\label{sec:experiments}

\begin{figure}[htbp]
    \centering
    \includegraphics[width=1\linewidth]{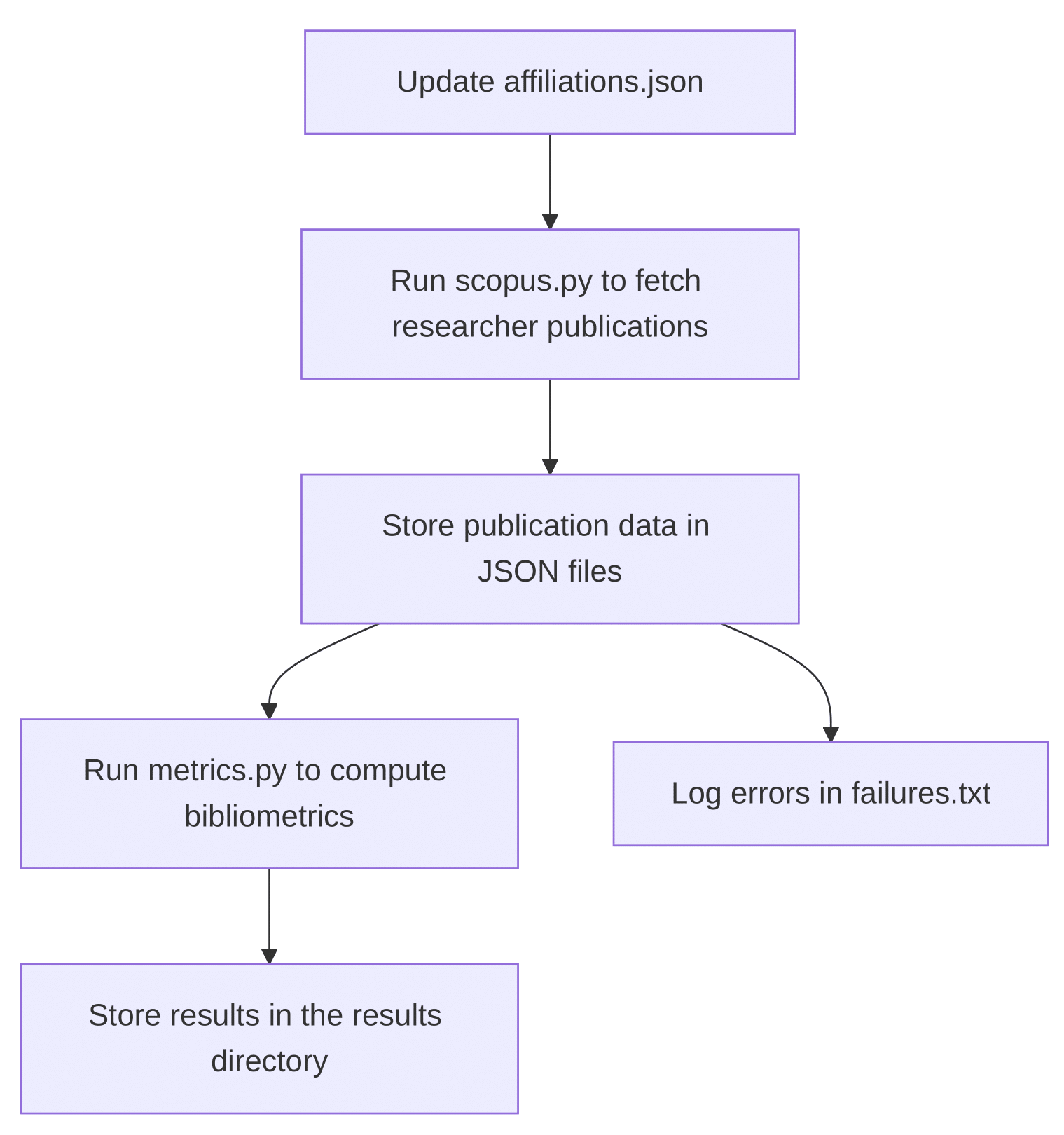}
    \caption{The flowchart shows the modules required for reusing/reproducing the results using data extracted from Scopus.}
    \label{fig:dataflow}
\end{figure}

The experimental process shown in Figure~\ref{fig:dataflow} consists of three main stages: managing researcher affiliations, retrieving publication data from the Scopus API, and calculating bibliometric indicators. Our setup is flexible, allowing for adaptation to different sets of researchers and institutions and provided in the code implementation on our  GitHub repository\footnote{\url{https://github.com/nepython/metrics.git}}.
We store the mapping between universities and their associated researchers in a configuration file, \texttt{affiliations.json}, located in the \texttt{data} directory. Each university is a key, with its corresponding researchers listed as values, and to adapt the data extraction for a different set of universities and researchers, one must modify this file by adding the relevant institution names and updating the associated list of researchers. An example structure of the \texttt{affiliations.json} file is as follows:

\begin{verbatim}
{
    "Institution Name": {
        "affiliation": "Institution Name",
        "scopus_id": [
            "Scopus ID 1", 
            "Scopus ID 2",
            "Scopus ID 3"
        ],
        "city": "City Name",
        "country": "Country Name",
        "researchers": {
            "Researcher Name 1": "...",
            "Researcher Name 2": "...",
            "Researcher Name 3": "..." 
        }
    },
    ... // more institutions
}
\end{verbatim}

After updating this file, the subsequent steps can be executed to fetch and process the new data set. The data retrieval process is covered in more detail in Section~\ref{subsec:data}. It is handled by the \texttt{scopus.py} script, which reads the \texttt{affiliations.json} file and accesses the Scopus API to retrieve each researcher's publication records. The script performs a search to retrieve the researcher's Scopus ID and publication details from the configuration file, including titles, author lists, citation counts, and journal information. The retrieved data is stored as JSON files in the \texttt{data/Scopus} directory for further analysis. To accommodate a different set of researchers or institutions, one may either modify the \texttt{affiliations.json} file or update the retrieval process by editing \texttt{scopus.py}. The script also handles API rate limiting by pausing when limits are reached and resuming when permitted.


Once publication data is retrieved, we employ the \texttt{metrics.py} script to compute various bibliometric indicators for each researcher. This script processes the JSON files produced during data retrieval and calculates metrics such as the h-index, hm-index, hl-index, and other customised indicators. We store the results in the \texttt{results} directory, where each researcher's computed bibliometrics are accessible for analysis.
The pipeline for metric computation is modular, allowing for the addition of new bibliometric measures as required. If errors arise during the data fetching or computation stages, they are logged in the \texttt{failures.txt} file, located in the \texttt{data} directory, for further inspection. For reusing this experimental setup for a different set of researchers and institutions, one needs to update the \texttt{affiliations.json} file, execute the \texttt{scopus.py} script to fetch the data, and then run the \texttt{metrics.py} script to compute the bibliometric indicators. The resulting data is then available for further analysis and reporting.

\section{Results}\label{sec:results}

\subsection{Data and bibliometrics}

We first analyse the publication data of the top-cited 400 Go8 Australian university researchers (50 per university), encompassing a total of 168,563 scholarly works. Among these, 154,337 (91.56\%) were published in journals, 8,517 (5.05\%) in conference proceedings, and 5,471 (3.25\%) as books or book series. Other venues included reports (0.004\%), trade journals (0.09\%), and unknown sources (0.05\%).

Table~\ref{tab:academic_metrics} presents a comparative summary of bibliometric metrics computed using this publication data. We observe a strong correlation between h-leadership and h-index, with the h-index decrementing slightly with the increase in the number of overall authors and a decrease in authorship contribution as measured by the authorship position. Notably, the h-leadership index consistently provides more granular differentiation compared to the h-index, particularly for researchers with high collaborative outputs. For instance, UNSW has the lowest median number of co-authors amongst papers analysed, which is also reflected in their h-leadership index (77) being close to h-index (80) post-weighted reduction. That being said, the number of co-authors isn't the sole criterion determining individual contribution. This is reflected in UniMelb's h-leadership index (74) remaining high compared to the h-index (78), despite its researchers having 11 median co-authors per paper. Comparatively, USyd, having a similar number of median co-authors per paper (12), scores 74 in the h-leadership index against its h-index of 86. This emphasises the fairer assessment enabled by the weighted citation scheme, which strikes a balance between the number of co-authors and the authorship position, not adversely penalising larger study groups.
 
 Table~\ref{tab:academic_metrics_individual} presents the hl-index for leading \textbf{Medicine} researchers defined by citations received on researchers' Google Scholar profiles in the selected Go8 Australian universities. For instance, Google Scholar provides a list of the top 10 researchers at UniMelb based on citations with links on the page to the next 10 researchers \footnote{\url{https://scholar.google.com/citations?view_op=view_org&hl=en&org=10194437184893062620}}. The reason for using Google Scholar for top author identification instead of relying on Scopus has been explained in Section \ref{subsec:data}. Similarly, results for \textbf{Engineering} discipline are shown in Table~\ref{tab:academic_metrics_individual_2} and data for all disciplines available in code repository. Discipline specific analysis is important as citation practices vary significantly between disciplines. We note that the top-cited researchers at Australian National University (ANU) and Monash University engage in Physics, Astronomy and Medicine disciplines   (Figure~\ref{fig:research_subject_areas_individual}), which explains the higher number of co-authors observed. We also note that Scopus search API clips a publication's authors to the first 100, and obtaining a full list requires using abstract retrieval API\footnote{\url{https://cran.r-project.org/web/packages/rscopus/vignettes/multi_author.html}}. Since an abstract retrieval API call needs to be made per publication, we decided against its usage due to existing weekly quota limits. We highlight that in our selected random samples, we have 2 author profiles with 100 median coauthors at ANU and 3 at Monash. We notice huge variations when comparing the h-index to our hl-index  (59 vs 4) for the 3rd ANU profile and the 5th ANU profile (80 vs 5). We also notice similar patterns when comparing the h-index and hl-index for Monash profiles (1st, 3rd, and 5th profiles. We provide a complete list of the top 50 profiles in \textit{metrics.csv} for all Go8 universities via our GitHub repository \footnote{\url{https://github.com/nepython/metrics/tree/main/results}}. We also notice that the first, single and last authorship makes a large impact on the hl-index when compared to the h-index, since authorship position is the key principle of the hl-index. In cases when a medium number of coauthors is low (2 or 3), then the hl-index is close to the h-index.

\begin{table*}
    \centering
    \begin{tabular}{lcccccc>{\centering\arraybackslash}p{2cm}>{\centering\arraybackslash}p{2cm}}
    \toprule
    University & Publications & Total citations & h-index & h-frac-index & hm-index & hl-index & Median number of co-authors & \% drop from h-index to hl-index \\
    \midrule
    ANU & 296 & 35001 & 68 & 18 & 23 & 53 & 25 & 22.06 \\
    UoA & 418 & 38887 & 81 & 19 & 28 & 69 & 25 & 14.81 \\
    Monash & 481 & 47674 & 75 & 14 & 25 & 64 & 18 & 14.67 \\
    UNSW & 450 & 41158 & 80 & 18 & 34 & 77 & 6 & 3.75 \\
    UQ & 442 & 52833 & 86 & 20 & 33 & 79 & 9 & 8.14 \\
    UWA & 348 & 36620 & 78 & 15 & 28 & 69 & 18 & 11.54 \\
    USyd & 552 & 45959 & 86 & 17 & 32 & 76 & 12 & 11.63 \\
    UniMelb & 376 & 38062 & 78 & 18 & 29 & 74 & 11 & 5.13 \\
    \bottomrule
    \end{tabular}
    \caption{Statistics on the number of publications, net citations, mean h-index, mean h-frac-index, mean hm-index, mean hl-index, mean of the median coauthors per researcher and comparison of hl with h for Go8 researchers based on Scopus data.}
    \label{tab:academic_metrics}
\end{table*}

\begin{table*}
\centering
\small
\begin{tabular}{p{3cm}>{\centering\arraybackslash}p{1.5cm}cccc>{\centering\arraybackslash}p{2cm}>{\centering\arraybackslash}p{2cm}l}
\toprule
Name & Median citations & h-index & hl-index & i10-index & hm-index & \% first, single, last authorship & Median number of co-authors & University \\
\midrule
Jenkins, Mark A. & 26 & 84 & 79 & 352 & 19 & 15.83 & 19 & UniMelb \\
Loi, Sherene M. & 36 & 102 & 101 & 238 & 19 & 36.18 & 15 & UniMelb \\
Herrman, Helen Edith & 12 & 49 & 49 & 157 & 19 & 46.69 & 5 & UniMelb \\
Wei, Andrew H. & 13 & 55 & 55 & 139 & 12 & 38.46 & 13 & UniMelb \\
Seymour, John Francis & 18 & 104 & 102 & 364 & 28 & 39.23 & 8 & UniMelb \\
\hline
Macaskill, Petra & 34 & 66 & 66 & 145 & 20 & 15.85 & 6 & USyd \\
Halliday, Glenda Margaret & 33 & 117 & 114 & 559 & 47 & 34.46 & 7 & USyd \\
Wang, Jiejin Jie J. & 38 & 120 & 116 & 586 & 44 & 28.45 & 7 & USyd \\
Bauman, Adrian Ernest & 21 & 121 & 121 & 735 & 54 & 40.49 & 6 & USyd \\
Young, Jane M. & 20 & 52 & 51 & 174 & 23 & 31.58 & 5 & USyd \\
\hline
Robinson, Bruce W.S. & 27 & 63 & 63 & 197 & 24 & 50.58 & 5 & UWA \\
Prince, Richard L. & 23 & 70 & 68 & 240 & 25 & 56.90 & 7 & UWA \\
Liu, Jianjun & 34 & 106 & 87 & 385 & 16 & 14.53 & 20 & UWA \\
Jablensky, Assen Verniaminov & 19 & 73 & 65 & 213 & 30 & 57.86 & 5 & UWA \\
Hankey, Graeme J. & 22 & 150 & 119 & 614 & 47 & 35.61 & 7 & UWA \\
\hline
Fitzgerald, Paul B. & 27 & 91 & 91 & 428 & 40 & 54.34 & 6 & ANU \\
Griffiths, Kathleen Margaret & 40 & 72 & 72 & 171 & 29 & 52.20 & 4 & ANU \\
Mann, Graham J. & 28 & 89 & 61 & 257 & 16 & 13.91 & 16 & ANU \\
Colquhoun, Samantha M. & 28 & 28 & 23 & 38 & 4 & 12.50 & 9 & ANU \\
Nolan, Christopher James & 17 & 37 & 36 & 66 & 13 & 35.65 & 6 & ANU \\
\hline
Norman, Robert John & 22 & 102 & 101 & 359 & 41 & 54.17 & 5 & UoA \\
Price, Timothy J. & 10 & 54 & 53 & 170 & 15 & 36.36 & 9 & UoA \\
Amare, Azmeraw T. & 55 & 58 & 46 & 88 & 4 & 13.56 & 100 & UoA \\
Munn, Zachary & 9 & 42 & 40 & 100 & 15 & 52.71 & 6 & UoA \\
Sanders, Prashanthan & 14 & 104 & 103 & 433 & 29 & 39.34 & 8 & UoA \\
\hline
Lalloo, Ratilal & 13 & 52 & 27 & 106 & 12 & 39.78 & 5 & UQ \\
Evans, David M. & 47 & 103 & 86 & 274 & 18 & 20.88 & 21 & UQ \\
Roberts, Jason A. & 15 & 87 & 86 & 411 & 33 & 47.38 & 7 & UQ \\
Montgomery, Grant W. & 39 & 158 & 120 & 673 & 30 & 16.14 & 16 & UQ \\
Charlson, Fiona J. & 240 & 55 & 40 & 68 & 5 & 26.74 & 19 & UQ \\
\hline
Cumpston, Miranda S. & 13 & 12 & 12 & 14 & 3 & 43.48 & 5 & Monash \\
Bailey, Michael J. & 19 & 95 & 95 & 444 & 33 & 12.21 & 7 & Monash \\
Li, Shanshan & 25 & 74 & 57 & 255 & 16 & 16.53 & 12 & Monash \\
McNeil, John J. & 13 & 79 & 78 & 398 & 31 & 39.85 & 7 & Monash \\
Cheng, Allen C. & 15 & 74 & 73 & 312 & 27 & 34.42 & 7 & Monash \\
\hline
Mattick, Richard P. & 21 & 62 & 62 & 198 & 25 & 49.64 & 6 & UNSW \\
Carr, Vaughan James & 24 & 72 & 67 & 226 & 25 & 38.60 & 7 & UNSW \\
Law, Matthew G. & 21 & 86 & 85 & 395 & 27 & 22.36 & 9 & UNSW \\
Brodaty, Henry & 20 & 105 & 101 & 574 & 42 & 40.67 & 7 & UNSW \\
Anstey, Kaarin J. & 23 & 83 & 83 & 391 & 41 & 54.53 & 5 & UNSW \\
\bottomrule
\end{tabular}
\label{tab:academic_metrics_individual}
\caption{h-leadership index for 5 randomly sampled top researchers in "Medicine" discipline at each Go8 institute. We derived the median citation number by sorting an author's publications by citation and considering the median publication's citation count. We sorted publications by the number of coauthors to obtain the median number of coauthors.}
\end{table*}

\begin{table*}
\centering
\small
\begin{tabular}{p{3cm}>{\centering\arraybackslash}p{1.5cm}cccc>{\centering\arraybackslash}p{2cm}>{\centering\arraybackslash}p{2cm}l}
\toprule
Name & Median citations & h-index & hl-index & i10-index & hm-index & \% first, single, last authorship & Median number of co-authors & University \\
\midrule
Yabsley, Bruce D. & 34 & 132 & 3 & 1109 & 10 & 0.21 & 100 & USyd \\
Eggleton, Benjamin J. & 1 & 97 & 96 & 477 & 42 & 66.98 & 6 & USyd \\
Driscoll, Tim Robert & 22 & 72 & 48 & 157 & 17 & 30.00 & 7 & USyd \\
Mai, Yiu Wing & 20 & 109 & 109 & 732 & 61 & 69.65 & 3 & USyd \\
Zomaya, Albert Y.H. & 6 & 79 & 79 & 487 & 40 & 75.09 & 4 & USyd \\
\hline
Colless, Matthew M. & 46 & 89 & 87 & 224 & 17 & 12.68 & 18 & ANU \\
Zheng, Liang & 31 & 51 & 51 & 85 & 18 & 45.59 & 4 & ANU \\
Hartley, Richard I. & 20 & 62 & 62 & 173 & 39 & 64.02 & 3 & ANU \\
Yin, Zongyou & 28 & 70 & 70 & 151 & 17 & 38.22 & 8 & ANU \\
McClelland, David E. & 28 & 105 & 36 & 355 & 14 & 21.05 & 100 & ANU \\
\hline
Hessel, Volker & 14 & 77 & 77 & 346 & 36 & 49.01 & 5 & UoA \\
Qiao, Shizhang Zhang & 84 & 175 & 175 & 492 & 58 & 71.56 & 6 & UoA \\
Govindan, Kannan & 39 & 110 & 110 & 326 & 55 & 49.53 & 4 & UoA \\
Jackson, Paul D. & 44 & 139 & 12 & 1199 & 12 & 0.89 & 100 & UoA \\
Jiao, Yan & 72 & 69 & 69 & 113 & 16 & 17.65 & 6 & UoA \\
\hline
Milburn, G. J. & 15 & 78 & 77 & 238 & 44 & 76.18 & 3 & UQ \\
Yamauchi, Yusuke & 33 & 147 & 147 & 945 & 59 & 51.83 & 8 & UQ \\
Keller, Jurg & 59 & 104 & 104 & 262 & 41 & 44.87 & 5 & UQ \\
Sawadsky, Andreas & 84 & 57 & 5 & 78 & 1 & 2.27 & 100 & UQ \\
Chen, Guanrong (Ron) & 15 & 146 & 146 & 1021 & 82 & 71.35 & 3 & UQ \\
\hline
Nash, Jordan A. & 33 & 141 & 27 & 1590 & 14 & 0.21 & 100 & Monash \\
Skands, Peter Zeiler & 20 & 38 & 31 & 63 & 14 & 64.76 & 4 & Monash \\
Maier, Stefan A. & 17 & 103 & 102 & 412 & 42 & 32.78 & 7 & Monash \\
Egede, Ulrik & 26 & 116 & 15 & 798 & 8 & 1.46 & 100 & Monash \\
Valencia, German & 15 & 38 & 34 & 112 & 22 & 85.71 & 2 & Monash \\
\hline
Fane, A. G.(Tony) & 44 & 119 & 119 & 471 & 62 & 55.59 & 4 & UNSW \\
Ng, Derrick Wing Kwan & 15 & 79 & 79 & 326 & 34 & 34.11 & 5 & UNSW \\
Dai, Liming & 19 & 166 & 166 & 733 & 69 & 48.90 & 5 & UNSW \\
Boufous, Soufiane & 25 & 46 & 33 & 98 & 12 & 30.66 & 8 & UNSW \\
Boyer, Cyrille Andre & 46 & 99 & 98 & 346 & 43 & 52.44 & 5 & UNSW \\
\bottomrule
\end{tabular}
\label{tab:academic_metrics_individual_2}
\caption{h-leadership index for 5 randomly sampled top researchers in "Engineering" discipline at six Go8 institutions. We derived the median citation number by sorting an author's publications by citation and considering the median publication's citation count. We sorted publications by the number of coauthors to obtain the median number of coauthors.}
\end{table*}

\subsection{Research collaboration trends}

We review the research collaboration trends across 8 Go8 Australian universities to examine what portion of the research is led by these universities, noting first, middle and last/senior authorship in published papers. In the scientific literature, the first, second and last authorship indicate research leadership while the middle authors indicate research collaboration. 

We find that the research output across Go8 is overwhelmingly multi-authored, with 5 out of the 8 institutions having less than 4\% single-authored papers and the remaining 3 being less than 7\%. Despite the majority of publications by Go8 researchers having them placed as middle-authored, the top researchers at 7 of the 8 institutes published at least 42\% of their output as single (~3\%), first (~30\%) or last (~9\%) authors, highlighting their significant contributions (Figure~\ref{fig:authorship_position}). UNSW leads the Go8 pack with the first and single-authored publications, followed by the University of Sydney(USyd). Surprisingly, the University of Melbourne (UniMelb) has shown to be the weakest when it comes to research leadership by first and single authors, although it ranked as the number 1 Australian university \footnote{\url{https://about.unimelb.edu.au/facts-and-figures}} by QS \footnote{QS Ranking: \url{https://www.topuniversities.com/university-rankings}} and Times of High Education (THE) \footnote{THE: \url{https://www.timeshighereducation.com/world-university-rankings}} ranking for almost a decade. 






Next, we present an analysis of our h-leadership index for assessing the research impact of leading researcher profiles and compare it with conventional metrics such as the h-index. We report  
Our findings quantitatively demonstrate the effectiveness of the h-leadership index (hl-index) in recognising the nuanced contributions of researchers within collaborative settings (Table~\ref{tab:academic_metrics}) when compared to established indexes such as the h-index, hm-index, h-frac index and citations. Our hl-index has been able to 
effectively penalise the h-index as the median number of co-authors grows.

\begin{figure}[htbp]
    \centering
    \includegraphics[width=1\linewidth]{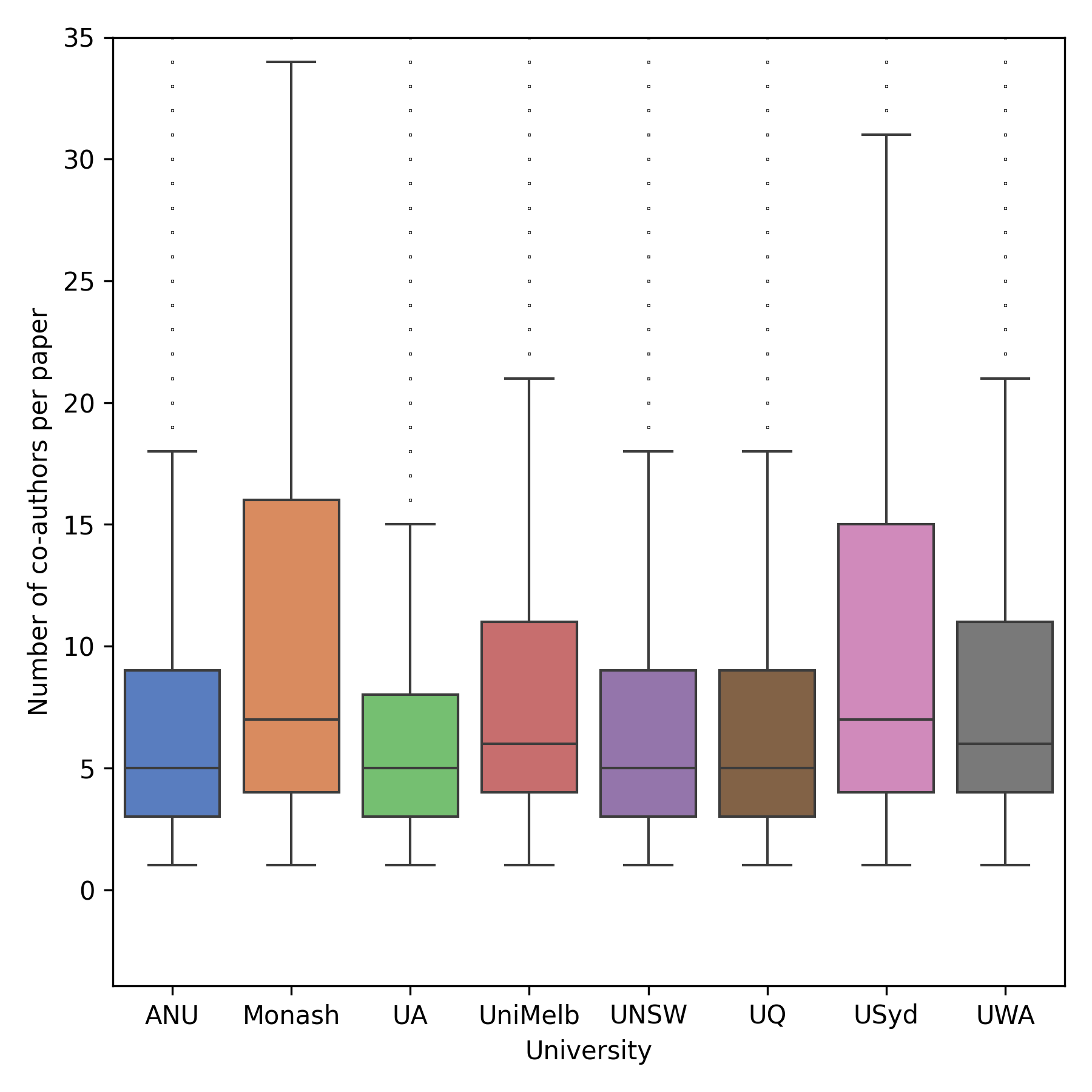}
    \caption{Distribution of co-authorship count for each publication by the Go8 researchers. The y-axis was clipped at 35 to better visualise the plots as outliers exceed 100 for each university.}
    \label{fig:coauthors}
\end{figure}

\begin{figure}[htbp]
    \centering
    \includegraphics[width=1\linewidth]{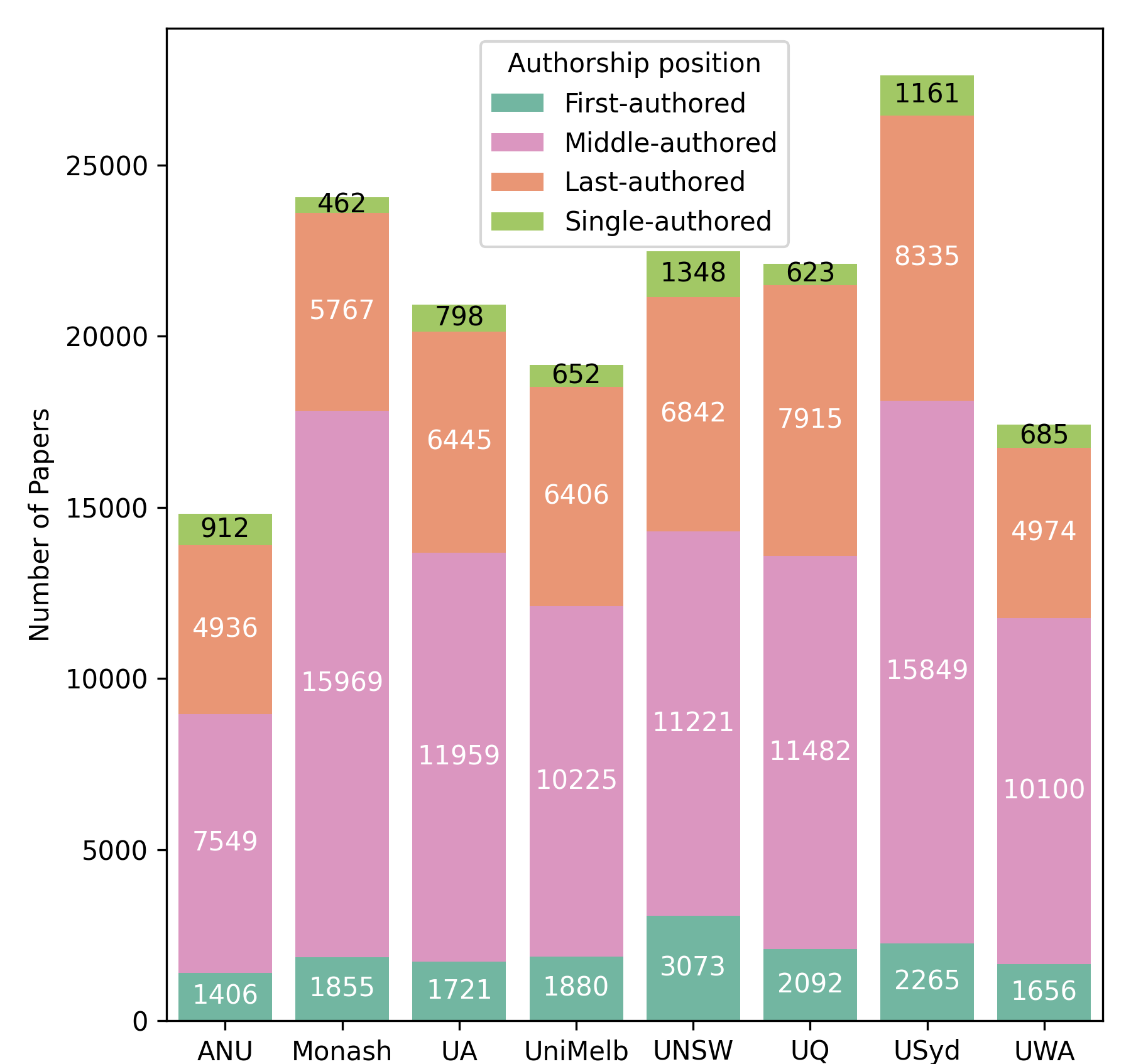}
    \caption{The authorship position number and percentages for the researchers at Go8. Most researchers' publications are middle-authored or last-authored. The single-authored numbers have not been included in either the first or last authored statistics.}
    \label{fig:authorship_position}
\end{figure}

We find that the authorship distribution across Go8 is mostly consistent with the median number of coauthors \footnote{This is different from the statistics in the table. We computed the median number of coauthors in Table~\ref{tab:academic_metrics}   by taking the average across each of the 50 author profiles, whereas the box plot concatenates all 50 author publications before plotting.} between 5 and 7 with Monash and the University of Sydney on the higher end (Figure~\ref{fig:coauthors}).

\subsection{Publication trends over time}

\begin{figure}[H]
    \centering
    \includegraphics[width=1\linewidth]{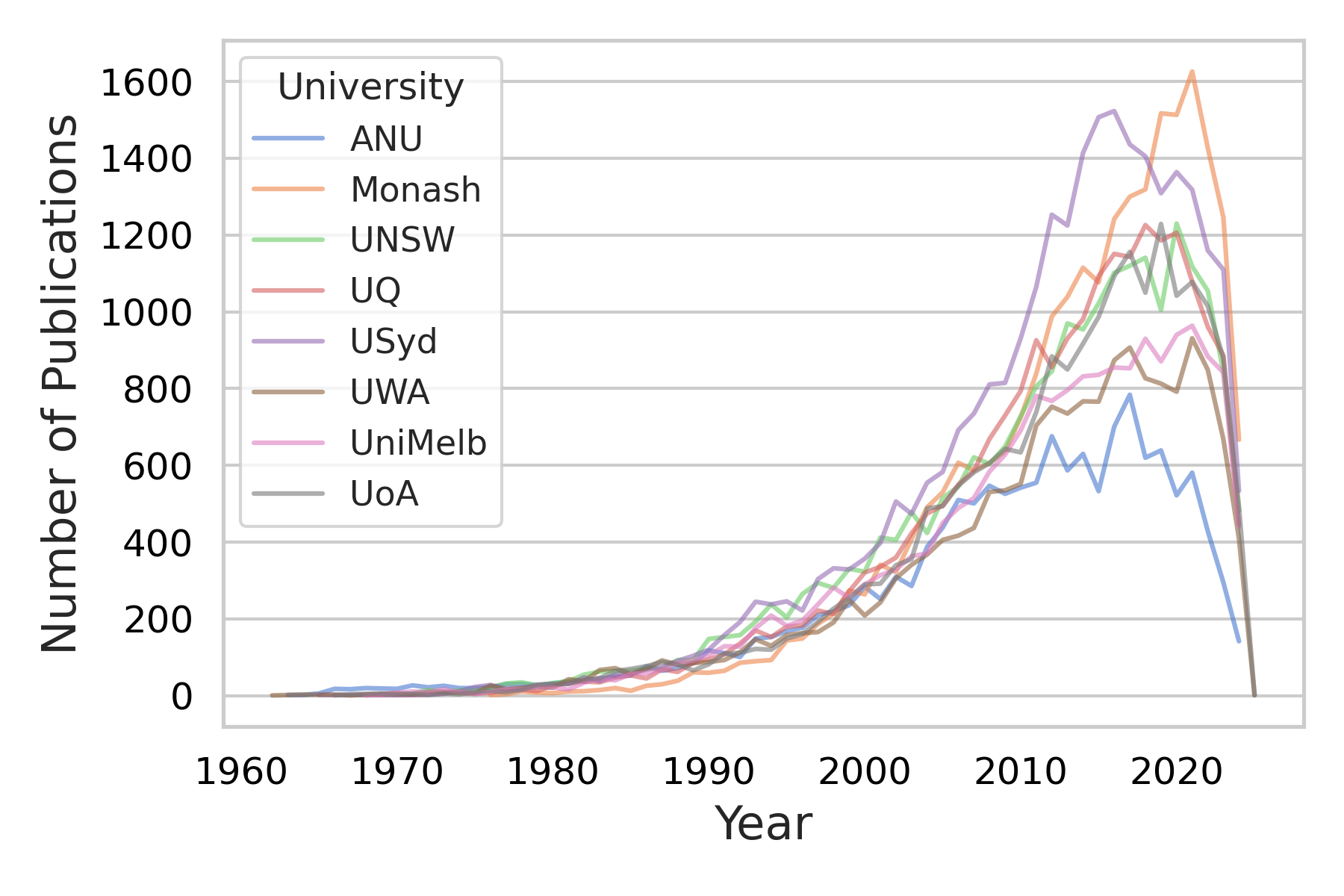}
    \caption{Temporal analysis of publication numbers by the Go8 researchers reveals a peak forming between 2018 and 2021, followed by a decline, possibly indicating that the Scopus data does not fully cover the most recently published articles.}
    \label{fig:temporal_analysis}
\end{figure}

\begin{figure}[H]
    \centering
    \includegraphics[width=1\linewidth]{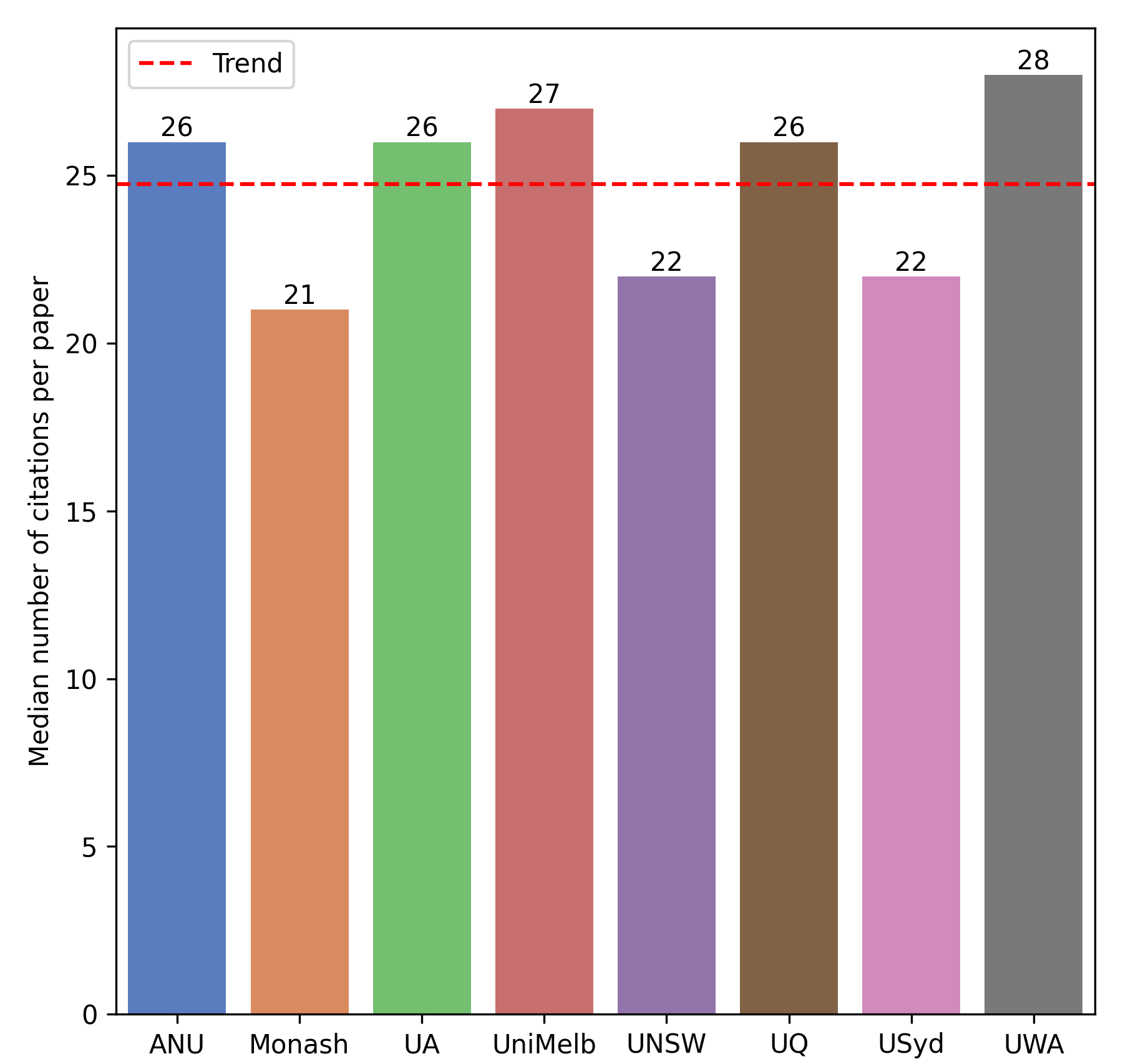}
    \caption{Median number of citations per publication for the top Go8 author profiles. This analysis highlights the research impact of each institution and provides insights into their contributions to academia.}
    \label{fig:go8_median_citations}
\end{figure}

We present the publication trend from January 1962 to December 2024 (Figure~\ref{fig:temporal_analysis} and observe that the highest number of articles was published around  2018-2021, possibly indicating that Scopus does not cover recent year publications adequately or that the top researchers have been reducing their publication output in recent years. We cannot determine the exact reason behind this behaviour. All universities also achieve a very high median number of citations greater than 20 per publication with values ranging between 21 to 28 as shown in Figure~\ref{fig:go8_median_citations}.

\subsection{Subject area distribution}

\begin{figure*}[htbp!]
    \centering
    \includegraphics[width=0.8\linewidth]{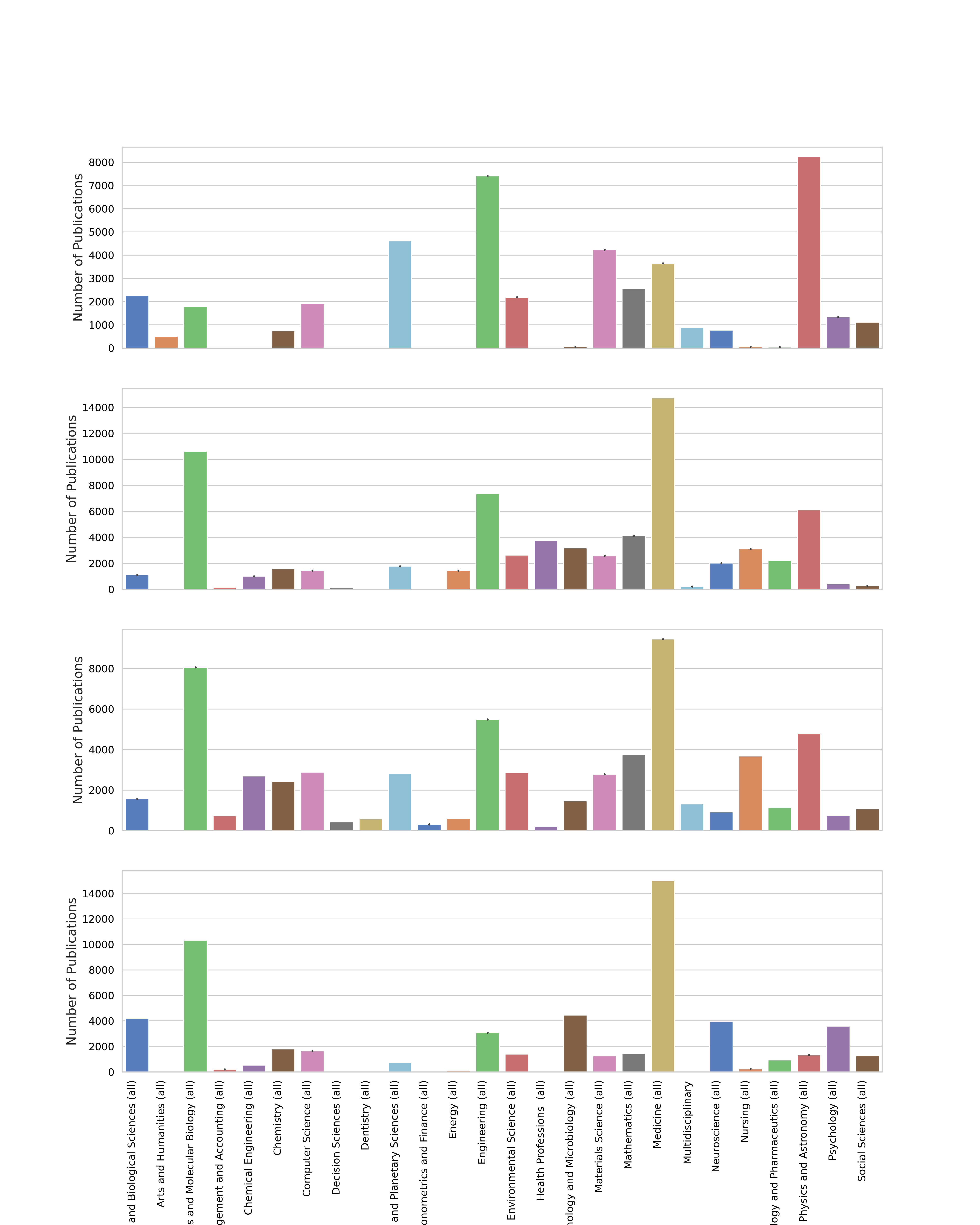}
    \caption{Research focus of Go8 top researchers by publication count: a) Australian National University, b) Monash University, c) University of Adelaide, d) University of Melbourne}
    \label{fig:research_subject_areas_individual}
\end{figure*}

We next present an analysis of the research impact for different subject areas of the top 50 cited researchers from Go8 Australian universities.  We note that the median number of citations received by the researchers per paper is 25 (Figure~\ref{fig:cites}) and the top 5 subject areas are: i) Medicine (98,785), ii) Biochemistry, Genetics and Molecular Biology (65,959), iii) Engineering (38,958), iv) Physics and Astronomy (33,289), v) Materials Sciences (24,792) (Table~\ref{tab:research_focus}). The Go8 universities perform well-balanced research across all disciplines, but when considering only the top 50 researchers, the distribution gets skewed in favour of specific disciplines. The top two subject areas based on the number of publications are given as follows. ANU leads in Physics, Astronomy and Engineering;  Monash, UA, UniMelb, UQ, USyd and UWA lead in Medicine and Biochemistry, Genetics and Molecular Sciences; UNSW leads in Medicine and Neuroscience (Figure~ \ref{fig:research_subject_areas_individual}). We find that Decision Sciences, Arts and Humanities, Economics, Econometrics and Finances, Dentistry and Business, Management and Accounting have significantly fewer publications than the other subject areas. This inconsistency in subject area distribution may impact the bibliometric analysis, explaining significant variation between universities.

\begin{table}
    \centering
    \small
    \begin{tabular}{lc}
        \toprule
        Subject area &                                             Publications \\
        \midrule
        Medicine &                                                 98,785 \\
        Biochemistry, Genetics and Molecular Biology &             65,959 \\
        Engineering &                                              38,958 \\
        Physics and Astronomy &                                    33,289 \\
        Materials Science &                                        24,792 \\
        Neuroscience &                                             24,193 \\
        Agricultural and Biological Sciences &                     22,938 \\
        Environmental Science &                                    22,087 \\
        Mathematics &                                              20,106 \\
        Chemistry &                                                17,184 \\
        Earth and Planetary Sciences &                             16,884 \\
        Immunology and Microbiology &                              16,856 \\
        Psychology &                                               16,567 \\
        Nursing &                                                  15,160 \\
        Social Sciences &                                          14,915 \\
        Computer Science &                                         12,298 \\
        Health Professions  &                                      10,251 \\
        Pharmacology, Toxicology and Pharmaceutics &                9,303 \\
        Chemical Engineering &                                      8,575 \\
        Energy &                                                    5,156 \\
        Multidisciplinary &                                         3,797 \\
        Business, Management and Accounting &                       2,489 \\
        Dentistry &                                                 1,514 \\
        Economics, Econometrics and Finance &                       1,503 \\
        Arts and Humanities &                                       1,419 \\
        Decision Sciences &                                           711 \\
        \bottomrule
    \end{tabular}
    \caption{Top research focus areas across Go8 Australian universities as per Scopus classification of publication output.}
    \label{tab:research_focus}
\end{table}

\begin{figure*}[htbp!]
    \centering
    \includegraphics[width=0.8\linewidth]{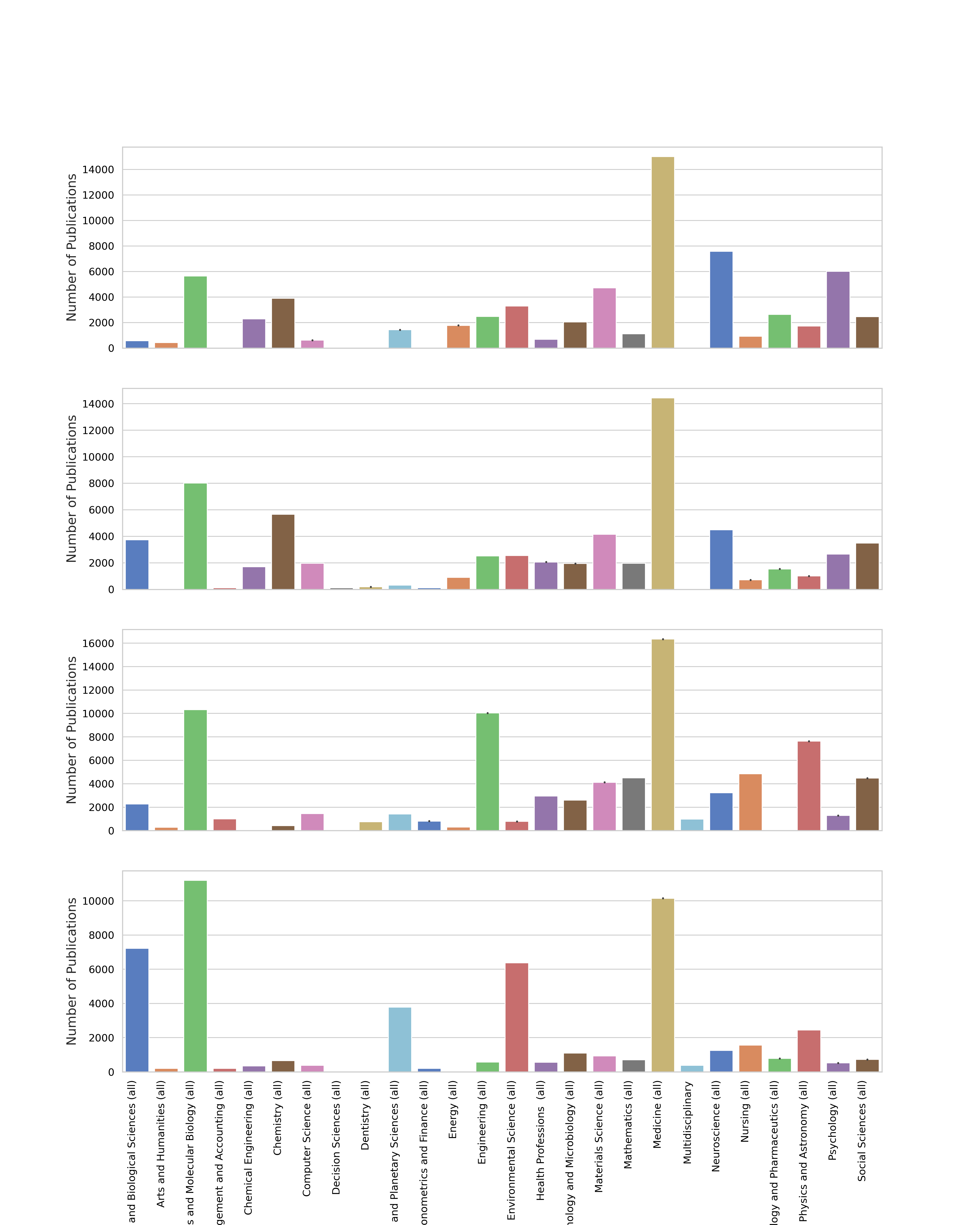}
    \caption{Research focus of Go8 top researchers by publication count: a) University of New South Wales, b) University of Queensland, c) University of Sydney, d) University of Western Australia.}
    \label{fig:research_subject_areas_individual_2}
\end{figure*}


\section{Discussion}
Our proposed hl-index introduced an innovative bibliometric approach, improving upon the key shortcomings of existing metrics such as the h-index, g-index, and c-score. We carried out a comparative analysis using the Scopus data for each of the metrics discussed so far, revealing some notable trends and demonstrating a strong research outlook across the Go8 Australian universities. We reviewed the collaboration patterns, publication impact, and research focus across universities. The weighting model in the hl-index adopts a complementary Gaussian distribution that optimally balances contributions (Figure~\ref{fig:h-leadership}), aligning well with established norms in authorship significance. Our analysis reveals that in large-scale collaborations, the h-leadership index highlights contributors more accurately, providing a reliable alternative to indices that neglect positional awareness.

This study also sheds light on collaboration trends within Go8 Australian universities.  The median number of co-authors (ranging from 6 to 25) underscored the collaborative nature of academic research in Australia (Table~\ref{tab:academic_metrics}). The hl-index captures these dynamics more effectively than traditional metrics, aligning with recent calls for greater inclusivity in research impact evaluation. The modular implementation of the hl-index allows for easy integration with other metrics, enabling holistic assessments. Universities and funding agencies can leverage the hl-index to evaluate contributions in team-based projects more equitably, fostering transparency in reward systems.

In the context of bibliometric analysis, several open-source tools have been developed to assist researchers in evaluating scholarly outputs. Bibliometrix ~\cite {dervics2019bibliometric}, an R package, offers comprehensive analysis and visualisation capabilities, allowing for a detailed exploration of bibliometric data. Similarly, ScientoPy ~\cite{ruiz2019software} is a Python package designed to handle large datasets and generate custom visualisations, making it accessible to those familiar with Python programming. Additionally, platforms like OpenAlex~\cite{priem2022openalex} provide open catalogues of scholarly papers, authors, and institutions, serving as valuable resources for bibliometric studies. Our Python hl-index module distinguishes itself by integrating the strengths of these existing tools while introducing unique features tailored to specific research, computing popular bibliometrics and visualising the authorship pattern, subject distribution, temporal characteristics, and citation trends. Our code is easily reproducible, extensible and modular, enhancing accessibility. Furthermore, by leveraging comprehensive data sources akin to OpenAlex, our toolbox ensures up-to-date and extensive bibliometric data coverage. This combination of versatility, user-centric design, and robust data integration positions your tool as a valuable asset for researchers seeking efficient and comprehensive bibliometric analysis solutions.


 We observe that the majority of publications are from life sciences, physical sciences, health sciences, and social science, which in comparison, had fewer publications.  Figure 9 presented differences in citation trends and research impact, eg, the impact factor of journals in the areas of medicine and computer science is generally much higher than arts and humanities.  In general, social sciences and humanities publications receive a lower number of citations and are not catered in the measurement of research impact \cite{reale2018review,bastow2014impact}. Aiello et al. \cite{aiello2021effective}  discussed strategies for better evaluation of the impact of social science and humanities research.
This is why metrics such as the Field Weight Citation Index (FWCI) \cite{purkayastha2019comparison} are becoming prominent with Scopus profiles providing this index for individual researchers and also institutions \cite{baas2020scopus}. Therefore, an institution that is heavily producing research papers in areas of science and technology will naturally have a higher impact in terms of citations when compared to an institution specialising in arts and humanities. Considering this, the Indian government has created a separate ranking system based on institution type (i.e. Universities and Technical Institutions) \cite{}, which needs to be considered for a better system. The ranking by THE and QS University Ranking \cite{ccakir2015comparative} does not take into account the nature of the institution in terms of areas of focus, the size of the university, and the funding received. The purpose of university ranking systems has been questioned \cite{vernon2018university} as it is natural to find institutions that have high funding with more research focus leading to higher ranking when compared to institutions that are focused on education delivery.  
Despite its advantages, the h-leadership index faces notable challenges:
\begin{enumerate}
    \item Sensitivity to the author listing order: The method assumes that authors were listed in order of their contribution, with the first and last authors being the most significant. Although this is used in science and engineering, it may not hold across all disciplines where authorship position conventions may differ.
    
    \item Discipline-specific citation practices: Variability in citation norms across disciplines can introduce biases, necessitating calibration of the weighting scheme for interdisciplinary use.
    
    \item Inclusion of Inactive Authors: The dataset included authors from all times, including inactive ones, potentially skewing the analysis.
    
    \item  Scopus data limitations: Discrepancies in author/coauthor affiliations and incomplete coverage of recent publications impact the accuracy of computed metrics. We further observed in our study that some authors' citation counts varied by an order of magnitude when cross-referenced with Google Scholar. The most recent years' publications also show limited coverage (Figure~\ref{fig:temporal_analysis}), which may be on account of Scopus considering only peer-reviewed journals, books and conference proceedings as sources. Preprints, datasets, software libraries and many other digital formats that aren't formally published but receive high citations thus get left out.
\end{enumerate}

The widespread adoption of the hl-index can be done by Scopus and Google Scholar profiles giving it as an additional metric. We note that the authorship collaboration trends are showing increased collaboration trends since quantity matters along with quantity for better citation trends. The prominence of Google Scholar's h-index and accessibility has motivated researchers from lower-ranked and low-income countries to feature a profile that has a way of comparing with researchers from higher-income universities/countries. However, over time, the h-index has been abused as there are cases of faculty leaders gaining coauthorship without contributions and mega-author publications where a single author publishes 70  papers in a year \cite{abduh2023unveiling,moreira2023rise,ioannidis2024evolving,price2018some}, known as hyper-prolific authors, which makes their contributions questionable. Furthermore, Ioannidis \cite{ioannidis2018thousands} demonstrated that thousands of prolific authors publish one paper every five days, which questions their actual contribution.

The c-score offered by Stanford's top 2 \% scientists list is a way forward \cite{ioannidis2019standardized}; however, it is an elitist approach where we find bibliometrics information only for the top researchers worldwide. This is demotivating for early and middle-career researchers, even in leading universities, as they fail to be listed since a fairly large number of citations in key papers are needed, which takes time to gather. The hl-index is a way forward, however, its implementation as a web resource and a service that is easily available can change the academic field. The use of the hl-index can then take place in university recruitment and promotions, which can eliminate the need to be mega-author publications in huge collaboration groups where there is little contribution.

Future work could explore the application of the h-leadership index across various disciplines and its potential integration with other metrics to further refine the evaluation of collaborative research efforts. Ultimately, the adoption of such advanced metrics will contribute to a more balanced and inclusive recognition of scholarly contributions in the academic community. Disciplines such as pure mathematics typically follow the author alphabetically, which is a limitation for the hl-index. In future works, the hl-index can be amended so that author profiles can collectively define their weightage in papers for those that are listed alphabetically. It would also be the case that a single author may be contributing to different fields with alphabetical author listing being used in some papers, while in others, contribution-based listing is used. An advanced version of the hl-index would take all these into account. Finally, the hl-index is just the beginning, and a lot can be done for better analysis of research impact.


\section{Conclusion}\label{sec:conclusion}

We presented the h-leadership index that provides a significant advancement in the evaluation of research performance, particularly in the context of multi-authored publications. We used a complementary Gaussian weighting scheme in the h-leadership index, which effectively acknowledges the varying levels of contribution among authors, offering a more nuanced and fair assessment than existing metrics. Our analysis of the top researchers across Go8 demonstrates the h-leadership index's ability to more accurately reflect the contributions of both leading and supporting authors. This metric not only enhances the recognition of first and last authors but also ensures that middle authors are fairly considered, albeit with appropriate weight adjustments.

The h-leadership index addresses key limitations in traditional bibliometric measures by integrating a positional awareness that was previously overlooked. As the landscape of academic publishing continues to evolve, metrics like the h-leadership index are essential for fostering a more equitable and comprehensive understanding of research impact. 

\section*{Code and Data Availability}

Open source code and data are available at our GitHub repository \url{https://github.com/nepython/metrics}.


 \bibliographystyle{elsarticle-num} 
 \bibliography{cas-refs}





\end{document}